\documentclass[prd,twocolumn,nofootinbib]{revtex4}
\usepackage{graphicx,epsfig}
\usepackage{epsfig}
\usepackage{mathrsfs}
\usepackage{bm}
\usepackage{verbatim}
\usepackage{amsfonts}
\usepackage[latin1]{inputenc}
\usepackage{graphicx}
\usepackage{amssymb}
\usepackage{bm}
\usepackage{color}
\usepackage{float}
\usepackage{amsmath}
\usepackage{amsfonts}
\usepackage{dcolumn}
\usepackage{bm}
\usepackage{hyperref}
\usepackage[normalem]{ulem}
\usepackage{mathtools}
\usepackage{cancel}

\def\rmG{{\rm G}}
\def\rmMpc{{\rm Mpc}}

\newcommand{\be}{\begin{equation}}
\newcommand{\ee}{\end{equation}}
\newcommand{\ba}{\begin{eqnarray}}
\newcommand{\ea}{\end{eqnarray}}

\begin{document}

\title{Cosmological chirality and magnetic fields from parity violating particle decays}
\author{Tanmay Vachaspati$^*$, Alexander Vilenkin$^{\dag}$}
\affiliation{
$^*$Physics Department, Arizona State University, Tempe, AZ 85287, USA. \\
$^\dag$Institute of Cosmology, Department of Physics and Astronomy, Tufts University, Medford, MA 02155, USA. \\
}

\begin{abstract}
\noindent
We estimate the chirality of the cosmological medium due to parity violating decays of standard 
model particles, focusing on the example of tau leptons. The non-trivial chirality is however too
small to make a significant contribution to the cosmological magnetic field via the chiral-magnetic
effect.
\end{abstract}

\maketitle

\section{Introduction}

The last few decades have seen growing interest in cosmic magnetic fields on several
fronts~\cite{Vachaspati:2020blt}. Several ideas have been proposed that can generate magnetic fields 
in cosmology, some of which are directly tied to known particle 
physics~\cite{Vachaspati:1991nm,Cornwall:1997ms,Vachaspati:2001nb} and its possible
extensions~\cite{Joyce:1997uy,Forbes:2000gr,Stevens:2012zz,Miniati_2018,Vachaspati:2020blt}.
The magneto-hydrodynamical (MHD) evolution of cosmological magnetic fields is now understood quite 
well on the basis of analytical arguments~\cite{Banerjee:2004df,Jedamzik:2010cy}
and direct simulations~\cite{Brandenburg:2017neh}.
There are claims for an indirect lower bound on the cosmological magnetic field 
strength~\cite{Neronov_Vovk_2010_science,Essey:2010nd,Dolag:2010ni,Wood:2017kcy,Biteau:2018tmv}, 
though not without debate~\cite{Broderick:2011av,Arlen:2012iy}, and more direct 
evidence~\cite{Chen:2014rsa}.
Concurrently there are claims of a parity violating signature that can be used to
measure the magnetic field helicity spectrum~\cite{Tashiro:2013ita,Chen:2014qva} though
with no significant detections as yet~\cite{asplund2020measurement,kachelriess2020searching}.

In parallel to these developments, motivated by heavy-ion collision 
experiments~\cite{Kharzeev:2011vv}, there has been renewed interest in chiral 
effects in plasmas, namely the chiral-magnetic \cite{Vilenkin:1980fu} and 
chiral-vortical \cite{Vilenkin:1979ui} effects (CME and CVE respectively). 
The CME and CVE have also been applied to the evolution of cosmological and
astrophysical magnetic fields
\cite{Joyce:1997uy,Boyarsky:2011uy,Tashiro:2012mf,Dvornikov:2012rk,Dvornikov:2013bca,
Dvornikov:2016jth,Brandenburg:2017rcb,Masada:2018swb}.
In this paper we discuss how CME and CVE can effectively arise 
in standard cosmology with standard particle interactions due to the parity-violating 
decays of heavy leptons and quarks. 
The basic idea is that the standard model has a number of unstable particles
that decay at various cosmological epochs, primarily due to the weak interactions.
Since the weak interactions violate parity, the decay products are chiral and
this provides a net particle helicity to the cosmological medium. 
The net particle helicity in principle leads to electric currents via the CME that 
can generate magnetic helicity. However, accounting only for decays of standard 
model particles, the net particle helicity is too small to significantly affect
cosmological magnetic fields and their helicity.

We start by describing the physical effect in some detail in the context of the tau lepton in 
Sec.~\ref{physics}, where we also estimate the induced electric currents. 
We find an upper bound to the magnetic helicity that can be generated due to
chiral effects in Sec.~\ref{maghel}.
Our conclusions are summarized and discussed in Sec.~\ref{conclusions}.

\section{Chirality production in tau decays}
\label{physics}

To illustrate the physics of the effect, in this section we will discuss the decay of tau leptons 
in the background of a magnetic field and fluid vorticity. Except for small differences, the 
physics carries over to the case of decays of other particles.

\subsection{Particle decay}
\label{particledecays}

Tau leptons decay into electrons (or muons) and neutrinos
\be
\tau^- \to e^- + \nu_\tau + {\bar \nu}_e
\label{taudec}
\ee
and anti-tau into positrons and neutrinos
\be
\tau^+ \to e^+ + {\bar \nu}_\tau + \nu_e
\label{antitaudec}
\ee
These decays violate parity since they proceed primarily by the weak interactions.
Therefore the tau predominantly decays into a relativistic left-handed electron, while 
an anti-tau decays into a relativistic right-handed positron. 
Due to the lepton asymmetry of the universe there are more taus than anti-taus,
and the cosmological medium gains net left-handed chirality as tau's decay.

The decay product electrons are chiral since they are produced by the weak
interactions, but chirality is not preserved for massive particles. Instead, as emphasized 
in Ref.~\cite{Grabowska:2014efa} in the context of supernovae and neutron stars, 
chirality is nearly equal to helicity for ultrarelativistic particles, so it is 
better to think of the final electrons as being in a definite helicity state. Helicity can only
change due to particle interactions. We shall adopt this view in what follows.

The $\tau$ mass is $m_\tau = 1777~{\rm MeV}$ and the $\tau$ lifetime in its rest frame is 
$\tau_\tau= 2.9\times 10^{-13}~{\rm s}$.  However, the decaying taus are constantly reproduced 
by reactions inverse to
(\ref{taudec}), (\ref{antitaudec}),\footnote{Tau-particles are also produced and destroyed in 
scattering reactions like  $\tau^- + {\nu}_e \to e^- + \nu_\tau$.  We disregard them in what follows, 
since they do not change the order of magnitude of the effect.} so the number density of taus, $n_\tau$, 
remains comparable to that of photons until the time
\be
t_\tau \sim 10^{-7}~{\rm s},
\label{ttaudecay}
\ee
when the cosmic temperature drops to $T\sim m_\tau$.  At later times $n_\tau$ decreases exponentially.

The particle helicity density, $n_\chi$, is produced in tau decays and is dissipated by 
helicity flipping scatterings and due to the chiral anomaly. The latter is proportional
to $\alpha^3 B^2$~\cite{Figueroa:2017hun}, where $\alpha \approx 1/137$ is the fine structure
constant and $B$ the magnetic field strength, and is much slower than helicity flipping
scatterings for vanishing or weak magnetic fields. We will ignore the anomalous flipping 
for now but will discuss it in Sec.~\ref{Bgen} when we consider the effect of particle
chirality on the generation of magnetic fields.
The evolution of particle helicity density can be described by the kinetic equation in the 
relaxation time approximation,
\be
\frac{d}{dt} (a^3 n_\chi) = 
\frac{a^3}{\tau_d} (\delta n_\tau - \delta n_\tau^{\rm eq}) - \frac{a^3 n_\chi}{\tau_\chi},
\label{1}
\ee
where 
\be
\delta n_\tau = n_\tau^+ - n_\tau^-,
\ee
$n_\tau^-$ and $n_\tau^+$ are the densities of tau and anti-tau particles, respectively, 
$\delta n_\tau^{\rm eq}$ is the equilibrium value of $\delta n_\tau$, 
$\tau_d  \sim (T/m_\tau)\tau_\tau$ 
is the decay time of taus (assuming that $T>m_\tau$ and with time dilation taken into 
account) and $\tau_\chi^{-1}$ is the electron 
helicity flipping rate.  At $T\gg m_e$, the helicity flipping rate 
is suppressed by a factor $m_e^2/T^2$ compared to the scattering rate 
$\alpha T$~\cite{Boyarsky:2020cyk} (earlier estimates of the scattering rate 
were suppressed by another factor of $\alpha$~\cite{Grabowska:2014efa}),
\be
\tau_\chi\sim \frac{1}{\alpha T} \frac{T^2}{m_e^2}.
\label{tauchi}
\ee

The excess of anti-tau's over tau's, $\delta n_\tau$, decreases due to tau decay and
is described by the equation,
\be
\frac{d}{dt} (a^3 \delta n_\tau)=\frac{a^3}{\tau_d} (\delta n_\tau^{\rm eq} - \delta n_\tau) .
\label{2}
\ee

At temperatures below the elecroweak phase transition, $T\lesssim T_{\rm EW}\sim 100$~GeV, we have 
$\tau_d \ll t$, where $t$ is the cosmic time\footnote{This is easily verified using the relation 
$t\sim m_{\rm P}/\sqrt{N} T^2$, where $m_{\rm P}$ is the Planck mass and $N$ is the number of particle species 
in equilibrium.}. 
This means that the equilibrium density of taus establishes very quickly (compared to the Hubble time), 
and the approximate solution of (\ref{2}) is 
$\delta n_\tau\approx \delta n_\tau^{\rm eq}$. Inserting \eqref{2} in \eqref{1} and then
using $\delta n_\tau\approx \delta n_\tau^{\rm eq}$ we have
\be
\frac{d}{dt} (a^3 n_\chi) = -\frac{d}{dt}\left(a^3 \delta n_\tau^{\rm eq}\right) - \frac{a^3 n_\chi}{\tau_\chi}.
\label{nchieq}
\ee
With a given $\delta n_\tau^{\rm eq}$, this equation can be solved in quadratures, but we shall instead find 
an approximate solution.  Since we are in the regime where $\tau_\chi \ll t$, the term on the left-hand side 
can be neglected and we obtain
\be
n_\chi\approx -\tau_\chi T^3\frac{d}{dt}\left(\frac{\delta n_\tau^{\rm eq}}{T^3}\right),
\label{nchi}
\ee
where we have used $aT\approx {\rm const}$.

Once we determine the equilibrium excess of anti-taus over taus, denoted by $\delta n_\tau^{\rm eq}$,
we can estimate the chirality density of the universe due to tau decays using \eqref{nchi}.

\subsection{Equilibrium density}

The equilibrium density $\delta n_\tau^{\rm eq}$ is given by
\be
\delta n_\tau^{\rm eq}=\frac{1}{2\pi^2}\int_0^\infty dp p^2 \left[f\left(\frac{E-\mu_\tau}{T}\right)
                                                                                               -f\left(\frac{E+\mu_\tau}{T}\right) \right],
\label{integral}
\ee
where $f(x)=(e^x +1)^{-1}$ is the Fermi distribution, $E=\sqrt{p^2 +m_\tau^2}$, and $\mu_\tau$ is the 
chemical potential of $\tau$ particles.  At $T\gg m_\tau, \mu_\tau$ we can expand the integrand in 
Eq.~(\ref{integral}) in powers of $m_\tau^2/p^2$ and $\mu_\tau/T$.  The integrations are then easily 
performed and we find
\be
\delta n_\tau^{\rm eq}\approx \frac{\mu_\tau T^2}{6} \left(1-\frac{3m_\tau^2}{2\pi^2 T^2}\right).
\label{3}
\ee

We assume that the baryon and/or lepton asymmetry of the universe was generated at $T\gg T_{EW}$ by 
some interactions beyond the Standard Model, for example by $(B-L)$-violating leptoquark decays.  This 
asymmetry was then redistributed between the Standard Model leptons and quarks by sphaleron processes, 
so at $T\ll T_{EW}$ we expect the chemical potentials of light baryons and leptons to be of the order 
$\mu/T\sim \eta_B$ ~\cite{Kuzmin:1987wn,Harvey:1990qw}, where $\eta_B \sim 10^{-9}$ is the observed 
baryon to photon ratio.  In the high-temperature regime, when $T$ is large compared to all 
relevant particle masses, we have $\mu_\tau /T\approx {\rm const}$, with a mass correction 
${\cal O}(m^2/T^2)$~\cite{Bochkarev:1989kp}.  Then Eq.~(\ref{3}) becomes
\be
\frac{\delta n_\tau^{\rm eq}}{T^3}\approx C\eta_B - K\eta_B\frac{m_\tau^2}{T^2},
 \label{33}
\ee
where $C$ and $K$ are ${\cal O}(1)$ numerical constants\footnote{This estimate assumes that 
taus are the heaviest particles present in equilibrium at temperature $T$. If a heavier particle is 
present in equilibrium, it too will contribute to the mass correction and may change the estimate.}.
The mass correction term in (\ref{33}) can be qualitatively understood as follows. 
As the temperature decreases, it becomes energetically favorable to transfer the conserved $\tau$-lepton 
number from $\tau$-particles to $\tau$-neutrinos.  The excess $\tau$-lepton number is also decreased as 
a result~\cite{Bochkarev:1989kp}.

Substituting Eq.~(\ref{33}) in (\ref{nchi}) we obtain
\be
n_\chi\approx -3K\eta_B \tau_\chi m_\tau^2 {\dot T}.
\ee
With ${\dot T}=-T/2t$, $t\sim m_{\rm P}/T^2$ and $\tau_\chi$ from Eq.~(\ref{tauchi}), this gives (omitting numerical factors)
\be
n_\chi \sim \frac{\eta_B m_\tau^2}{\alpha m_e^2}\frac{T}{m_{\rm P}} n_\gamma ,
\label{nchi2}
\ee
where $n_\gamma\sim T^3$ is the photon number density.

This estimate was derived assuming $T\gg m_\tau$, but it still applies at $T\sim m_\tau$.  
Reactions (\ref{taudec}), (\ref{antitaudec}) remain in equilibrium when $T$ drops well below 
$m_\tau$.  In this regime, $\delta n_\tau$ and $n_\chi$ decrease exponentially.

Similar formulae can be written down for the decay of other unstable particles. The largest
helicity is injected by the decay of the heaviest particle into the lightest particle and at
the highest temperature.

\section{Magnetic helicity}
\label{maghel}

As noted in Ref.~\cite{Brandenburg:2017rcb},
the maximum magnetic helicity that can be obtained from particle helicity can be
derived from the chiral anomaly equation, which can be written as a conservation
law,
\be
n_\chi + \frac{4\pi}{\alpha} h = {\rm constant}.
\ee
where $h = \langle {\bf A}\cdot {\bf B} \rangle$ is the magnetic helicity. Assuming 
that the initial magnetic helicity and the final particle helicity vanish, we get
\be
h_{\rm max} = \frac{\alpha}{4\pi} n_\chi
\sim \frac{\eta_B m_\tau^2}{4\pi m_e^2}\frac{T}{m_{\rm P}} n_\gamma 
\ee
where we have used \eqref{nchi2}. We compare $h_{\rm max}$ to magnetic 
helicity that could be induced due to baryogenesis~\cite{Cornwall:1997ms,Vachaspati:2001nb},
\be
h_B \sim \frac{\eta_B}{\alpha} n_\gamma \sim 10^{-5} \, {\rm cm}^{-3}
\sim 10^{-45}\, \rmG^2\, \rmMpc
\ee
where we have used the known cosmic baryon number density and are
using natural units.
Then
\be
h_{\rm max} \sim \frac{\alpha m_\tau^2}{4\pi m_e^2}\frac{T}{m_{\rm P}} h_B
\sim 10^{-10} h_B
\ee
where we have used $T \sim 100\, {\rm GeV}$ in the numerical estimate. Even if the
decay of top quarks with mass $\sim 175\, {\rm GeV}$ to down quarks with mass
$\sim 1\, {\rm MeV}$ is considered, $h_{\rm max} \sim 10^{-6} h_B$. Comparing
to observations, even with the most conservative lower bound of $10^{-19}\, \rmG$
on Mpc scales, we get an estimate for the observed helicity $\sim 10^{-38}\, \rmG^2\, \rmMpc$.

\section{Conclusions}
\label{conclusions}

We have shown that the decays of certain {\it standard model} particles can lead to a chiral
cosmological medium around the epoch of the electroweak phase transition. The final
result for the chiral asymmetry due to tau-lepton decays is given in \eqref{nchi2}. However,
the asymmetry is suppressed by the baryon to entropy ratio ($\eta_B \sim 10^{-9}$) and
the effect on magnetic field helicity generation is very weak as we have shown in
Sec.~\ref{maghel}. Nonetheless it is of interest that the cosmological medium 
was chiral at the earliest moments even within the standard model of particle physics.

\section{Acknowledgements}

We thank the participants of the Nordita workshop on ``Quantum Anomalies
and Chiral Magnetic Phenomena'', especially Axel Brandenburg and Kohei Kamada
for feedback. We also thank Matt Baumgart, Cecilia Lunardini, Igor Shovkovy, and 
Tracy Slatyer for discussions. TV's work is supported by the U.S. Department of Energy, 
Office of High Energy Physics, under Award No.~DE-SC0019470 at Arizona State 
University. AV is supported by the National Science Foundation Award No.~1820872.

\bibstyle{aps}
\bibliography{paper}

\end{document}